\documentclass[twocolumn,showpacs,preprintnumbers,amsmath,amssymb]{revtex4}
\usepackage{graphicx}% Include figure files
\usepackage{dcolumn}% Align table columns on decimal point
\usepackage{bm}% bold math
\begin{document}

\title{Resonant inelastic x-ray scattering (RIXS) spectra of magnesium 
diboride
}

\author{K.\ Kokko$^{*}$, V.\ Kulmala, and J.A.\ Leiro}
\affiliation{Department of Physics, University of Turku, FIN-20014 Turku, 
Finland} 
\author{W.\ Hergert} 
\affiliation{Department of Physics, Martin-Luther-University, 
Friedemann-Bach-Platz 6, D-06099 Halle, Germany}

\date{\today}

\begin{abstract} 
Using the tight-binding linear muffin-tin orbitals method, the soft x-ray 
fluorescence K-emission spectra of boron in MgB$_2$, excited close to the 
absorption edge, are estimated. In the calculations the angle of incidence 
$\theta$ 
between the direction of the incoming photon and the hexagonal axis of the 
specimen is $60^{\rm o}$ and $75^{\rm o}$. 
Comparison with experiment is possible in the former case where good 
agreement is found. 
Furthermore, another resonant feature 
below the Fermi energy is predicted for the larger angle. This feature 
can be related to the excitations to the antibonding B $\pi$-band in the 
neighbourhood of the L--H line in the Brillouin zone.
\end{abstract}
\maketitle

Magnesium diboride is a new kind of superconductor having a critical 
temperature of 39 K \cite{Akimitsu01}. Although it was earlier widely 
used in chemical 
technology, the superconducting property of this compound was not discovered 
until now \cite{Nagamatsu01}. At the moment, MgB$_2$ has attracted 
considerable attention 
regarding spectroscopic investigations \cite{Vasquez01}. For instance, angle 
resolved
photoemission from single crystals have revealed clear dispersions of the 
occupied valence bands \cite{Uchiyama02}. Additional information has been 
obtained using 
x-ray absorption - (XAS) and emission spectroscopy (XES) \cite{Kurmaev02}. 
In x-ray 
K-absorption a core-hole is created for the 1s level by a photon and the 
resulting photoelectron is escaping from the specimen. The K-emission band can 
be measured when the deep lying level will be filled by a valence-electron 
and another photon is emitted. The resonant inelastic scattering (RIXS) 
differs from the conventional XES in the sense that the photoelectron 
remains in the conduction band above the Fermi surface and it will be 
absorbed by the sample \cite{Kotani01}. 

Utilizing synchrotron radiation, resonant inelastic x-ray scattering 
is a promising method to probe element-specific, local 
momentum-resolved electronic structure of systems that are difficult to 
investigate using other techniques. The symmetry of the  
occupied and  unoccupied states is coupled with the 
polarization and direction of the incoming and outgoing radiation. 
Crystal momentum conservation has been observed in many materials 
e.g.\ in hexagonal boron nitride and graphite \cite{Carlisle99}. 
Theoretical consideration of the RIXS within the band structure 
picture is given e.g.\ in \cite{Ma94}.

In RIXS, a photon ($\hbar\omega_{q}$) comes to the sample, 
excites a core level ($\epsilon_a$) electron to the conduction state 
($\epsilon_c$) and a valence electron ($\epsilon_v$) 
drops to the core hole emitting a photon ($\hbar\omega_{q'}$). 
The formula for the doubly differential cross section in the dipole 
approximation (the wave vectors of the ingoing and outgoing photons are small, 
$q \approx q' \approx 0$, compared to 
the dimensions of the Brillouin zone of the Bloch states ($\vec{k}$\,) of the 
electrons) implemented in our x-ray spectrum program (XSPEC) 
\cite{Hergert} is
\begin{eqnarray}
\frac{d^2\sigma}{d\Omega d(\hbar\omega_{q'})}\propto
&&\frac{\omega_{q'}}{\omega_q}\sum_{b_c,b_v,\vec{k},m_s}\left|\sum_{t,m_j}
M^{b_v,\vec{k}}_{t,m_j,m_s}M^{* b_c,\vec{k}}_{t,m_j,m_s}\right|^2
\nonumber\\ 
&&\times\delta (\epsilon_a-\epsilon_c+\hbar\omega_q)
\delta(\epsilon_a-\epsilon_v+\hbar\omega_{q'}),
\end{eqnarray}
where $b_c$, $b_v$, $m_s$, $m_j$ and $t$ are indexes for conduction and 
valence bands, spin, magnetic quantum number and an 
atom in the unit cell. Here $t$ summation includes those atoms in the unit 
cell which have the core state of the specified energy $\epsilon_a$. Using the 
conventional 
LMTO notation \cite{Skriver84} the matrix element has the form \cite{Laihia98}
\begin{equation} 
M^{b,\vec{k}}_{t,m_j,m_s}=\sum_{l,m}i^l(A^{b,\vec{k}}_{t,l,m}M^r_{t,l}
+B^{b,\vec{k}}_{t,l,m}\dot{M}^r_{t,l})M^a_{m_j,l,m,m_s},
\end{equation}
where $l$ and $m$ are angular momentum and magnetic quantum numbers of the 
valence and conduction states. $M^r_{t,l}$, 
$\dot{M}^r_{t,l}$ and $M^a_{m_j,l,m,m_s}$ are radial matrix 
element, its energy derivative and angular matrix element, respectively, 
having the following properties. 
\begin{equation}
M^{\rm{r}}_{t,l}=
\int R_{t}(r)r
\phi_{t,l}(r)r^2 dr
\end{equation}
\begin{equation}
\dot{M}^{\rm{r}}_{t,l}=
\int R_{t}(r)r
\dot{\phi}_{t,l}(r)r^2 dr
\end{equation}
\begin{equation}
M^{\rm{a}}_{m_j,l,m,m_{\rm{s}}}=
<j,m_j|\hat{\epsilon}\cdot\hat{r}|l,m,m_{\rm{s}}>,
\end{equation}
where $R_{t}(r)$ is the radial part of the core-level wave function 
and $\phi_{t,l}(r)$ is that of the valence or conduction band, 
$\hat{\epsilon}$ is the polarization vector of the photon, 
$j$ refers to the total angular momentum of the core state and 
$<j,m_j|$ and $|l,m,m_{\rm{s}}>$ are the angular parts of the core and 
valence/conduction states, respectively. 
The angular matrix element leads to the selection rules 
between the initial and final electronic states in the absorption 
and emission of a photon in the 
scattering process ($\Delta l=\pm1$). 
The localized core electron state of an atom $t$ centered at a point 
$\vec{r}_{\rm a}$ is
\begin{equation}
\phi^{\rm{a}}_{t,m_j}(\vec{r}\,)
=R_{t}(|\vec{r}-\vec{r_{\rm{a}}}|)|j,m_j>.
\end{equation}
$A_{t,l,m}^{b\vec{k}}$ and $B_{t,l,m}^{b\vec{k}}$ are coefficients 
which depend on 
the crystal structure and the potential of the investigated system. The 
wave function of the valence and conduction electrons has the form 
\begin{equation}
\Psi^{b\vec{k}}(\vec{r}\,) = \sum_{t,l,m}i^{l}
(A_{t,l,m}^{b\vec{k}}\phi_{tl}(r)+
B_{t,l,m}^{b\vec{k}}\dot{\phi}_{tl}(r))|l,m,m_{\rm s}>,
\end{equation}
where $\phi_{tl}(r)$ and $\dot{\phi}_{tl}(r)$ are the partial 
wave and its energy derivative. 
LMTO wave functions are linearized with respect to energy at some suitable 
fixed energy $E_{\nu l}$. Because we are interested in energy eigenvalues 
$E_{b\vec{k}}$ near to $E_{\nu l}$ we can substitute $B_{t,l,m}^{b\vec{k}}$ 
coefficients by their approximate form 
$A_{t,l,m}^{b\vec{k}}(E_{b\vec{k}}-E_{\nu l})$, which reduces the required 
memory in computations about a factor of two.

The electronic structure calculations were performed using the 
scalar-relativistic tight-binding linear muffin-tin orbital method in the 
atomic sphere approximation (ASA) \cite{Andersen85}. The valence states 
consisted of Mg 3$s$, 3$p$ and 3$d$ states and B 2$s$, 2$p$ and 3$d$ states. 
The hexagonal unit cell contained three atoms and three empty spheres. 
We used lattice parameters $a=b=2.99$ ${\rm \AA}$ and $c=3.41$ 
${\rm \AA}$ corresponding to equilibrium volume in our calculations. 
For the exchange-correlation potential 
the parametrized form by Perdew and Zunger \cite{Perdew81} 
was used. The number of $\vec{k}$ points was 648 in the whole Brillouin zone. 

The geometry used in the calculations is the following. The direction of the 
absorbed photon makes an angle $\theta$ with respect to the hexagonal 
$c$-axis of the MgB$_2$ single crystal. The absorbed photon is linearly 
polarized and the polarization is in the plane containing the direction of the 
photon and the $c$-axis of the crystal. In this way the component of the 
polarization vector of the absorbed photon along the $c$-axis can be varied 
from 0 to 1 by increasing $\theta$ from 0$^{\rm o}$ to 90$^{\rm o}$. 
The direction of the emitted photon is perpendicular to the direction of 
the absorbed photon and makes a 90$^{\rm o}-\theta$ angle with 
respect to the $c$-axis. For the emitted photon we have used 
two different linear polarizations. The polarization is either in the 
($a,b$)-plane or along the direction of the absorbed photon. All our 
results are expressed using an energy scale where the Fermi energy is 0 eV.

\begin{figure}
\includegraphics{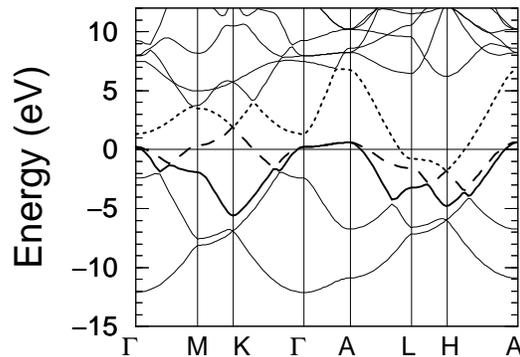}
\caption{\label{fig1} 
Band structure of MgB$_2$. Bands 3, 4 and 5, which cross the Fermi 
level, are shown by a thicker line: solid, dashed and dotted, respectively.}
\end{figure}

Overall features of both XES and XAS of MgB$_2$ are well produced 
by band structure calculations \cite{Nakamura01}. Thus RIXS experiments 
are also expected to be explained within the ordinary band structure 
picture. 
Because the valence hole is delocalized and the 1s core hole in boron is well 
screened \cite{Kokko02} it is reasonable to use ground state orbitals to 
obtain 
theoretical spectra. Due to the conservation of both energy and 
momentum we need detailed band structure data in order to interpret 
the spectra. The calculated band structure is shown in Figure~\ref{fig1}. 
Since we 
are interested in absorption and emission of photons coupled with 
electronic states near to the Fermi energy, the most relevant bands are 
those which cross the Fermi level, i.e.\ bands number 3--5. Zhang {\em et al.} 
\cite{Zhang03} have measured RIXS spectra at different angles of 
incidence ($\theta$ = 15$^{\rm o}$, 45$^{\rm o}$, 60$^{\rm o}$ with respect to 
the $c$-axis of the MgB$_2$ crystal) and using different excitation energies 
(187.25 eV -- 188.25 eV, Fermi energy: $E_F$ = 187.28 eV). They found two 
peaks in their spectra. The intensity of the peak just below the 
Fermi energy depends clearly on 
both direction and energy of the exciting radiation. On the other hand, the 
intensity of the peak at about 2 eV below the Fermi energy was 
considered to be unchanging. 

\begin{figure}
\includegraphics{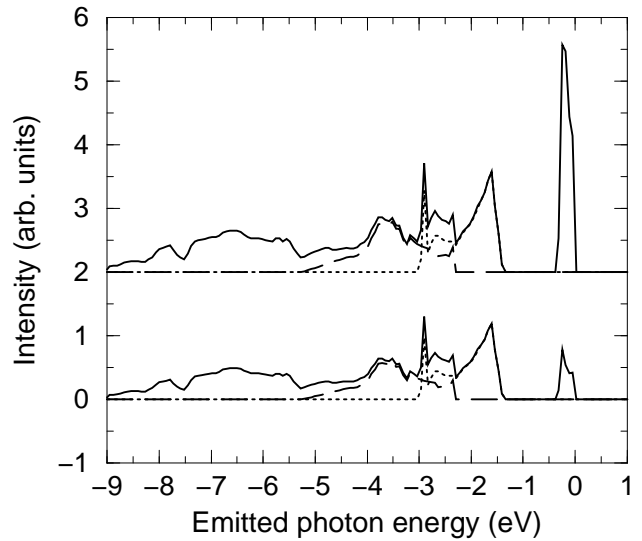}
\caption{\label{fig2} 
RIXS spectra calculated at $\theta=60^{\rm o}$. Excitation energy 
is 0.2 eV relative to the Fermi energy. Polarization of the emitted 
photon is in the ($a,b$)-plane (upper spectrum) and in the direction of the 
absorbed photon (lower spectrum). Dashed and dotted lines correspond to 
spectra calculated using bands (3,5) and (4,5), respectively.}
\end{figure}
In Figure~\ref{fig2} we show our results for the $\theta=60^{\rm o}$ 
incidence in the 
case of two polarizations of the emitted photon. 
The unpolarized emission 
spectrum is an average of these two extreme cases. The peak just below the 
Fermi energy is solely due to absorption to the band 4 and emission from the 
band 3. The features between $-4$ eV and $-1.5$ eV consist almost entirely 
from RIXS involving band pairs (3,5) and (4,5). 
By increasing angle $\theta$ one can measure more directly excitations to 
the $\pi$ band. 

\begin{figure}
\includegraphics{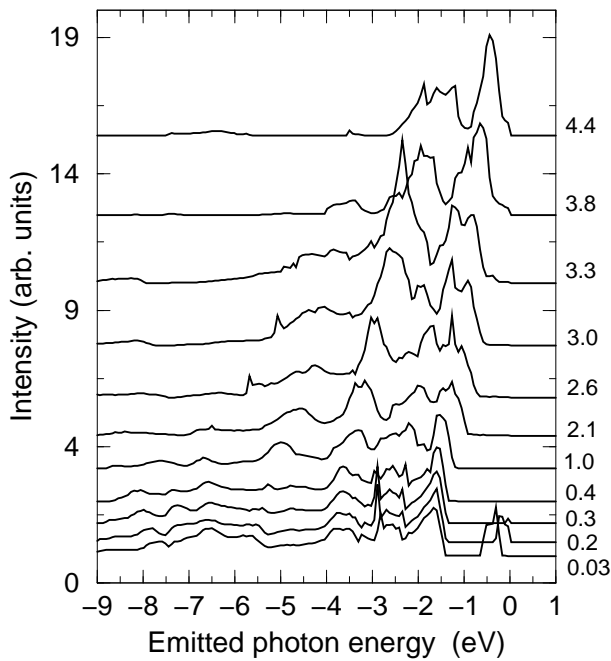}
\caption{\label{fig3} 
RIXS spectra calculated at $\theta=75^{\rm o}$. Exitation energy 
(in eV), 
relative to the Fermi energy, is shown beside each spectrum. Polarization 
of the emitted photon is in the ($a,b$)-plane.}
\end{figure}

\begin{figure}
\includegraphics{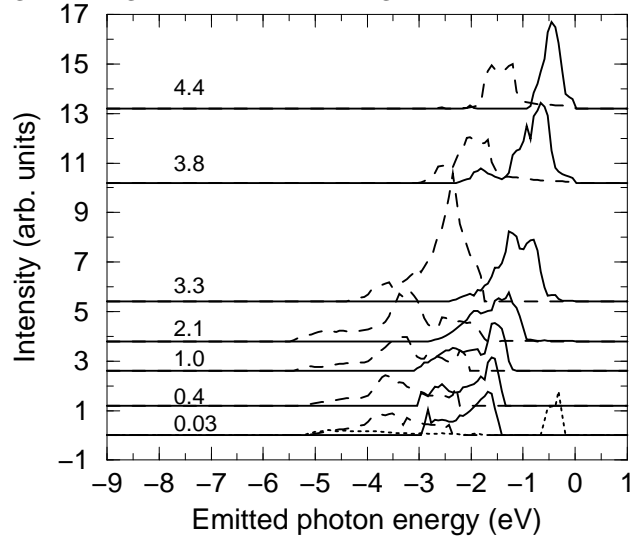}
\caption{\label{fig4} 
Some of the spectra shown in Fig. 3 calculated using the bands 
(3,4) (dotted), (3,5) (dashed) and (4,5) (solid).}
\end{figure}

In Figure~\ref{fig3} we show the spectra corresponding to 
$\theta=75^{\rm o}$ 
and the polarization vector of the emitted radiation, in our calculation, 
being in the $ab$-plane. 
Near the top of the valence band 
there appears a peak which disappears when the excitation 
reaches 0.4 eV above the Fermi energy. The broad structure from $-4$ eV to 
$-1.5$ eV shows not much dispersion up to excitation energies 0.4 eV above 
the Fermi energy. However, using excitation energies above 1 eV the 
intensity 
of the peaks increases considerably and the position of the peaks shifts 
about 1 eV closer to 
the Fermi level as the excitation energy increases. To decide what 
bands and what parts of the Brillouin zone 
are responsible for each peak in the spectrum we have calculated some of the 
spectra band by band. As Figure~\ref{fig4} shows the peak around $-2$ eV (low 
excitation energies) is mainly due to transitions involving bands 4 and 5. 
In the same way, the peak around $-3.5$ eV is mainly due to transitions 
involving bands 3 and 5. From Figure~\ref{fig1} we see that both of these 
peaks 
are produced close to the L--H line in the Brillouin zone. In this region, 
near the Fermi energy, the constant energy surfaces of the antibonding 
$\pi$ band have a tubular form \cite{Kortus01}. 

In summary, we have calculated RIXS spectra of MgB$_2$ using 
Kramers-Heisenberg formula with TB-LMTO 
eigenvalues and eigenstates for occupied and unoccupied electronic states.  
The results presented show that MgB$_2$ can be 
analysed and interpreted on the basis of ground state band structure 
calculations. In addition we predict nonlinear emission structures related 
with the excitation of 1s electrons to the unoccupied antibonding $\pi$ band 
in the neighbourhood of the L--H line in the Brillouin zone.

\vspace{5mm}
We acknowledge computer resources of CSC -- Scientific Computing Ltd., Espoo, 
Finland. This work has been in part supported by the Academy of Finland, 
Grant No.\ 51583 (K.K.\ and V.K.) and the Turku University Foundation 
(K.K.\ and W.H.). 

\vspace{5mm}
\noindent $^{*}$Author to whom correspondence should be addressed. Electronic 
address: kokko@utu.f{i}


\begin{thebibliography}{99}
%\bibitem{}
\bibitem{Akimitsu01}J. Akimitsu, Symposium on Transition Metal Oxides, Sendai, 
January 10, 2001.
\bibitem{Nagamatsu01} J. Nagamatsu, N. Nakagawa, T. Muranaka, Y. Zenitani 
and J. Akimitsu, Nature (London) 410, 63 (2001).
\bibitem{Vasquez01} R.P. Vasquez, C.U. Jung, Min-Seok Park, Heon-Jung Kim, 
J.Y. Kim, and Sung-Ik Lee, Phys. Rev.B, 64, 052510 (2001).
\bibitem{Uchiyama02} H. Uchiyama, K.M. Shen, S. Lee, A. Damascelli, D.H. Lu, 
D.L. Feng, Z.-X. Shen, and S. Tajima, Phys. Rev. Lett. 88, 157002 (2002).
\bibitem{Kurmaev02} E.Z. Kurmaev, I.I. Lyakhovskaya, J. Kortus, A. Moewes,
N. Miyata, M. Demeter, M. Neumann, M. Yanagihara, M. Watanabe, T. Muranaka, 
and J. Akimitsu, Phys Rev. B 65, 134509 (2002).
\bibitem{Kotani01} A. Kotani and S. Shin, Rev. Mod. Phys. 73, 203 (2001). 
\bibitem{Carlisle99} J. A. Carlisle, E. L. Shirley, L. J. Terminello, 
J. J. Jia, T. A. Callcott, D. L. Ederer, R. C. C. Perera, and F. J. Himpsel, 
Phys. Rev. B {\bf 59}, 7433 (1999).
\bibitem{Ma94} Y. Ma, Phys. Rev. B {\bf 49}, 5799 (1994); P. D. Johnson
and Y. Ma, Phys. Rev. B {\bf 49}, 5024.
\bibitem{Hergert} W. Hergert, K. Kokko, and R. Laihia (unpublished).
\bibitem{Skriver84} H.\ L.\ Skriver, {\it The LMTO Method} 
edited by M.\ Cardona and P.\ Fulde, Springer 
Series in Solid-State Sciences Vol. 41 
(Springer, Berlin, 1984).
\bibitem{Laihia98} R. Laihia, K. Kokko, W. Hergert, and J.A. Leiro, 
Phys. Rev. B {\bf 58}, 1272 (1998).
\bibitem{Andersen85} O.\ K.\ Andersen, O.\ Jepsen, and D.\ Gl\"otzel, 
{\it Proc.\ Int.\ School of Physics 'Enrico Fermi' (Course LXXXIX) 
Highlights of Condensed-Matter Theory}, 
edited by F.\ Bassani, F.\ Fumi, and M.\ P.\ Tosi, 
(North-Holland, Amsterdam, 1985), p.\ 59. 
\bibitem{Perdew81} J. P. Perdew and A. Zunger, Phys. Rev. B {\bf 23}, 
5048 (1981). 
\bibitem{Nakamura01} J. Nakamura, N. Yamada, K. Kuroki, T.A. Callcott, 
D.L. Ederer, J.D. Denlinger, and R.C.C. Perera, Phys. Rev. B {\bf 64}, 
174504/1-4 (2001).
\bibitem{Kokko02} K. Kokko, V. Kulmala, and J. A. Leiro, Phys. Rev. B 
{\bf 66}, 165114 (2002) 
\bibitem{Zhang03} G.P. Zhang, G.S. Chang, T.A. Callcott, D.L. Ederer, 
W.N. Kang, Eun-Mi Choi, Hyeong-Jin Kim, and Sung-Ik Lee, 
cond-mat/0302466
\bibitem{Kortus01} J. Kortus, I.I. Mazin, K.D. Belashchenko, V.P. Antropov, 
and L.L. Boyer, Phys. Rev. Lett. {\bf 86}, 4656 (2001). See Fig.\ 3 
in their article, where the antibonding $p_z$ band gives the red tubular 
part of the Fermi surface around the L--H line.  
\end{thebibliography}
\end{document}